\documentclass[twocolumn]{article}

\usepackage[colorlinks = true, citecolor = blue]{hyperref}
\usepackage{graphicx} 
\usepackage{xcolor}
\usepackage{amsmath, amssymb}
\usepackage{amsthm}
\usepackage{cleveref}
\usepackage[algo2e, inoutnumbered, algoruled, vlined]{algorithm2e}
\usepackage[utf8]{inputenc}
\usepackage{bbold}
\usepackage{algorithm}
\usepackage{comment}
\usepackage{thm-restate}
\usepackage{multirow}
\usepackage{titlesec} 
\usepackage{circuitikz}
\usepackage{algorithmicx}
\usepackage{algpseudocode}
\usepackage{diagbox}
\usepackage{tikz}

\usepackage[round]{natbib}
\newtheorem{definition}{Definition}

\title{\texttt{CriteoPrivateAds}: A Real-World Bidding Dataset to Design Private Advertising Systems}
\author{Mehdi Sebbar$^{1}$, Corentin Odic$^{1}$, Mathieu Léchine$^{1}$, Aloïs Bissuel$^{2}$, Nicolas Chrysanthos$^{1}$,\\ Anthony D'Amato$^{1}$, Alexandre Gilotte$^{1}$, Fabian Höring$^{2}$, Sarah Nogueira$^{1}$ and Maxime Vono$^{1}$}
\date{%
    $^1$Criteo AI Lab, Paris, France; 
    $^2$Criteo, Paris, France\\[2ex]%
}

\begin{document}

\maketitle

\begin{abstract}
    In the past years, many proposals have emerged in order to address online advertising use-cases without access to third-party cookies. 
    All these proposals leverage some privacy-enhancing technologies such as aggregation or differential privacy.
    Yet, no public and rich-enough ground truth is currently available to assess the relevancy of aforementioned private advertising frameworks.
    We are releasing the largest, in terms of number of features, bidding dataset specifically built in alignment with the design of major browser vendors proposals such as Chrome Privacy Sandbox.
    This dataset, coined \texttt{CriteoPrivateAds}, stands for an anonymised version of Criteo production logs and provides sufficient data to learn bidding models commonly used in online advertising under many privacy constraints (delayed reports, display and user-level differential privacy, user signal quantisation or aggregated reports).
    We ensured that this dataset, while being anonymised, is able to provide offline results close to production performance of adtech companies including Criteo - making it a relevant ground truth to design private advertising systems.
    The dataset is available in Hugging Face: \url{https://huggingface.co/datasets/criteo/CriteoPrivateAd}.
\end{abstract}

\section{Introduction}

\noindent \textbf{Context.} Over the past decade, consumer privacy concerns \citep{10.5555/1881151.1881153} and legislation have led browser vendors (e.g., Safari, Mozilla Firefox or Google Chrome) to restrict the possibilities to gather user information across the open internet, notably via the (future) deprecation of third-party cookies.
This operational constraint initiated a deep mutation of the online advertising ecosystem, which remains central to the funding of a large part of the open internet. 
The first major change was in 2017 when Apple introduced Intelligent Tracking Prevention (ITP) in Safari, limiting third-party cookies and cross-site tracking \citep{ITP_Safari}. 
Apple also strengthened privacy with App Tracking Transparency (ATT) in April 2021, requiring user consent for cross-app tracking on iOS \citep{10.1145/3531146.3533116}.
Mozilla followed in 2018 with Enhanced Tracking Protection (ETP) enabled by default in Firefox, while privacy-focused browsers like Brave implemented stricter measures \citep{ETP_Mozilla}.
In early 2020, Google announced it would phase out third-party cookies in Chrome due to privacy concerns. 
In contrast to other browser vendors, Google Chrome made the choice to replace the direct access to cross-domain user signals by a suite of application programming interfaces (APIs) called the Privacy Sandbox, enabling marketers and advertising companies to continue addressing advertising use-cases \citep{PSB,PSB_kleber}.
The Privacy Sandbox proposal has been the result of multiple discussions across the industry, merging ideas from several actors such as SPARROW from Criteo, TERN, PARROT or Dovekey.
Initially planned for 2022, third-party cookie deprecation on Chrome has been postponed several times due to regulatory and industry concerns.
In July 2024, after a Market Testing phase, Google revealed a user choice mechanism in Chrome \citep{PSB_user_choice}, allowing users to enable or disable third-party cookies, reflecting the tension between privacy protection and business models reliant on personalised advertising.
In 2024, Microsoft Edge also revamped its early private advertising proposal called MaskedLARk \citep{maskedlark} and published Ad Selection API sharing the same API backbone as Google Chrome Privacy Sandbox.
In the meantime, discussions involving advertisers, publishers, and technologists are also happening in global consortiums, such as the W3C, with the long-term objective of setting up common browser vendors API standards regarding private advertising on the open internet. 
Among those discussions, two main contributions focused on attribution and measurement have emerged, namely Privacy-Preserving Attribution (PPA) originated from several proposals from Apple, Meta and Mozilla Firefox; and Cookie Monster from Columbia University and Meta researchers \citep{10.1145/3694715.3695965}.
As evidenced through all aforementioned proposals, the whole advertising industry is seeking alternatives that balance privacy with the economic value of targeted advertising. 
A key area of investigation is leveraging privacy-enhancing technologies, such as $k$-anonymity, differential privacy, secure multi-party computation or trusted execution environments, to mitigate privacy risks while preserving advertising functionalities. 
As an example, Chrome Protected Audience and Aggregated Reporting Attribution APIs are using differential privacy in order to answer queries on user data.\\

\noindent \textbf{Contributions.} Albeit many private advertising proposals have emerged, the advertising industry is currently lacking a common ground truth in order to assess the privacy/utility trade-off of such envisioned private advertising systems. 
A public dataset, using synthetic or anonymised industrial data, would allow benchmarking of privacy-preserving mechanisms, fostering transparency, innovation, and ensuring effective solutions for user data protection and online advertising use-cases.
This paper is an attempt to fill this gap.
We are open-sourcing \texttt{CriteoPrivateAds}: the largest real-world anonymised bidding dataset in terms of number of features from Criteo production data.
Leveraging this dataset, researchers and engineers would be able to:
\begin{itemize}
    \item assess the drop of performance associated to the removal of cross-domain user signals and hence illustrating one of the impacts of third-party cookie deprecation on adtech companies;
    \item design and test private bidding optimisation approaches leveraging both non-sensitive contextual signals and user features to predict click and conversion events;
    \item design and test the relevancy of answers provided by aggregation APIs for measurement and learning bidding models.
\end{itemize}

\section{Related Work}
\label{sec:related_work}

\texttt{CriteoPrivateAds} is versatile enough to be leveraged to assess the offline performance of both non-private and private advertising approaches. 
Regarding the latter, benchmarks of bidding models based on differential privacy (DP), or other obfuscation techniques such as aggregation and signal quantisation (e.g. restricting user signals to a few bits) are of relevant interest, being both studied in the academic and industrial research and engineering communities.
Since DP stands for the workhorse methodology to derive private training frameworks, we will focus primarily on DP and show how this open-source release is necessary to benchmark private advertising systems with respect to current open-source bidding datasets.\\

\begin{table}[ht]
\centering
\renewcommand{\arraystretch}{1.2} 
\footnotesize
\begin{tabular}{|p{2cm}|p{1cm}|p{3.3cm}|}
\hline
\textbf{Data} & \textbf{Rows} & \textbf{Features} \\ \hline 
KDD Cup Track 2 \citep{kddcup2012_track2} & 5M & Search engine context, with queryn ad features, user id and click information.\\ \hline
Criteo-Kaggle Display Advertising Challenge \citep{criteo-display-ad-challenge} 
& 100K  &  Over 7 days of live traffic, 39 features hashed and fully undisclosed, click label.  \\ \hline
Avazu dataset \citep{Avazu2015} & 40M & Over 11 days of live traffic, 9 anonymised features, 6 contextual features, 5 user device features, click label. \\ \hline
Criteo 1TB Click Logs dataset \citep{criteo_1tb_dataset}
 & 4B  & Over 7 days of live traffic, 39 features hashed and fully undisclosed, click label. Similar to the dataset in Criteo-Kaggle-Display challenge. \\ \hline
 Criteo Attribution Modeling for Bidding Dataset \citep{DiemertMeynet2017}
 & 16.5M  & Over 30 days of live traffic, 9 contextual features, attribution data, user and campaigns ids, click and conversion labels. \\ \hline
\end{tabular}
\caption{Public Datasets for Bidding.}
\label{tab:dataset_overview}
\end{table}

\noindent \textbf{Differential Privacy and APIs.} Differential privacy (DP) has gained popularity over the last two decades due to its strong mathematical properties and privacy guarantees \citep{dwork2006calibrating,dwork2014algorithmic}. 
As DP is envisioned to be implemented at scale in ad measurement systems to replace third-party cookies \citep{ghazi2024differentialprivacyinteractivityprivacy}, it may become one of the largest real-world deployment of DP, impacting daily queries and users. 
Local differential privacy (LDP) adds noise to individual data points before they are sent to a statistician or report collector. 
While this avoids dependence on a trusted entity, LDP tends to degrade utility due to the large amount of noise required \citep{duchi2013privacyawarelearning,10.1007/978-981-96-0576-7_9}. Within the Privacy Sandbox context, the Attribution Reporting API (ARA) supplies locally DP-ed event-level reports.  
On the other hand, global differential privacy (GDP) operates by aggregating data across multiple users before noise is applied, which minimises the individual contributions and then lowers the noise. Though, it relies on a trusted aggregator who collects and aggregates data from multiple sources and ensures that the privacy guarantees are maintained throughout the process. 
As an illustration, the Privacy Sandbox by Google Chrome introduces the Private Aggregation Service that provides aggregated and globally DP-ed reports \citep{psb_paa}. 
Recently, Chrome proposed another path, albeit still in research phase, to learn bidding models by using differentially private stochastic gradient descent (DP-SGD), a GDP variant of SGD \citep{Abadi_2016}. 
It adds Gaussian noise to the gradients during training, which are clipped to control their range and then the privacy budget. This algorithm demonstrates good empirical performance, while incurring a reasonable privacy cost \citep{chua2024scalable,DBLP:conf/adkdd/DenisonGK0MNSVZ23}. Though, in the Privacy Sandbox setting, running DP-SGD necessitates access to a trusted server, for instance leveraging an execution environment (TEE), for the computation of gradients, their aggregation, the addition of noise and the management of privacy accounting.\\

\noindent \textbf{Bidding Public Datasets and Competitions.}
Advertising click-through-rate (CTR) prediction models have been assessed using various benchmarks \citep{Yang_2022}. 
The most frequently used public datasets are summarised in \Cref{tab:dataset_overview}.
Unfortunately, none of them can be used to evaluate the impact of private model training on CTR prediction based on browser vendors APIs constraints. 
Indeed, those datasets do not include sufficient information about features types (e.g., contextual, cross-domain, single-domain), availability at inference time and associated sensitivity from browser vendors lens, limiting their use for relevant benchmarking.
In complement to open-source datasets, some public competitions and challenges have been organised in order to assess the utility of private bidding model training methodologies, such as using noisy label proportions \citep{criteo-display-ad-challenge,fan2014learning,quadrianto2008estimating}.

\section{Background on Differential Privacy} 

As outlined in \Cref{sec:related_work}, DP stands for the most common tool to devise private machine learning training methodologies, including in the advertising industry.
For the sake of completeness and as we will provide DP-based private bidding model training frameworks in \Cref{sec:use-cases}, we recall hereafter some background and definitions on DP.\\

\noindent We consider $\mathcal{M}$ a \textit{randomised} algorithm that maps a dataset $\mathcal{D}$ to an output $S \in \text{Range}(\mathcal{\mathcal{M}})$. Two datasets $\mathcal{D}$ and $\mathcal{D}'$ are said to be neighbors and denoted $\mathcal{D} \sim \mathcal{D}'$, if they are the same up to one row. 

\begin{definition}
    \textbf{$(\varepsilon,\delta)$-DP}. Let $\varepsilon >0, \delta \geq 0$. 
    A mapping $\mathcal{M}$ satisfies $(\varepsilon,\delta)$-DP if for any $\mathcal{D} \sim \mathcal{D}'$ and any $S \in \text{Range}(\mathcal{M})$, 
    $$
    \mathbb{P}(\mathcal{M}(\mathcal{D}) \in S) \leq e^{\varepsilon} \mathbb{P}(\mathcal{M}(\mathcal{D}')) + \delta \ .
    $$ 
    The special case of $\delta=0$ is referred to as $\varepsilon$-DP.
\end{definition}

\begin{definition}
\label{def:laplace-dp}
\textbf{Laplace Mechanism}. Let $f$ be a mapping with range $\mathbb{R}^d$. The Laplace mechanism with parameter $b >0$ stands for a specific randomised algorithm $\mathcal{M}$ which outputs $f(\mathcal{D}) + \eta_{\text{L}}$, where $\eta_{\text{L}} \in \mathbb{R}^d$ stands for a zero-mean Laplace random variable with scale $b$. 
For any $\varepsilon>0$, the Laplace mechanism is $\varepsilon$-DP if $b = \Delta_1(f)/\varepsilon$,
where $\Delta_1(f) \! = \! \underset{\mathcal{D} \sim \mathcal{D}'}{\operatorname{sup}} \| f(\mathcal{D}) \! - \! f(\mathcal{D}') \|_1$. 
\end{definition}
For the sake of illustration, the Laplace mechanism defined in \Cref{def:laplace-dp} is currently used in the Privacy Sandbox Aggregated Reporting Attribution API to output noisy aggregated reports.

\begin{definition}\textbf{Gaussian Mechanism}. Let $f$ be a mapping with range $\mathbb{R}^d$. The Gaussian mechanism with parameter $\sigma >0$ stands for a specific randomised algorithm $\mathcal{M}$ which outputs $f(\mathcal{D}) + \eta_{\text{G}}$, where $\eta_{\text{G}} \in \mathbb{R}^d$ stands for a zero-mean Gaussian random variable with standard deviation $\sigma$. 
For any $\varepsilon, \delta>0$, the Gaussian mechanism is $( \varepsilon, \delta)$-DP if $\sigma = \Delta_2(f)\sqrt{2\log(1.25/\delta)}/\varepsilon$, 
where $\Delta_2(f) \! = \! \underset{\mathcal{D} \sim \mathcal{D}'}{\operatorname{sup}} \| f(\mathcal{D}) \! - \! f(\mathcal{D}') \|_2$. 
\end{definition}

\begin{definition}
\label{def:random-response}
    \textbf{Randomised Response}. Let \( f \) be a mapping with range \( E \) such that \( K = \text{Card}(E) < \infty \). For a given row $r \in \mathcal{D}$ and a given $p\in[0,1]$, let 
\[
\widetilde{\mathcal{M}}(r) = \begin{cases} 
f(r) & \text{with probability } p , \\
\mathrm{Uniform}(E) & \text{with probability } 1-p \ .
\end{cases}
\]
The randomised mechanism $\widetilde{\mathcal{M}}$ is $\varepsilon$-DP if $p(\varepsilon) = e^\varepsilon/(e^\varepsilon + K - 1)$.
\end{definition} 
For the sake of illustration, the randomised response mechanism defined in \Cref{def:random-response} is currently used in the Privacy Sandbox Protected Audiences API to output noisy 12-bit single-domain user features.

\begin{definition}\textbf{Label DP}. In a machine learning context, label differential privacy stands for a specific instance of differential privacy where only the labels are considered sensitive and hence need to be protected with differential privacy. 
The features vectors are considered not sensitive \citep{NEURIPS2021_e3a54649}. 
\end{definition}

\begin{definition}\textbf{User-level DP}. User-level differential privacy stands for a specific instance of so-called item-level differential privacy where two datasets $\mathcal{D}$ and $\mathcal{D}'$ are neighboring if at most one of the user’s contributions differ.
In this setting, a natural way to quantify a user contributions is to consider all rows in $\mathcal{D}$ and $\mathcal{D}'$ that refer to this user \citep{10.5555/3540261.3541215}. 
\end{definition}

\section{\texttt{CriteoPrivateAds} Dataset} \label{sec:3}

This section describes how the bidding dataset \texttt{CriteoPrivateAds} has been built, including business risk mitigations that were taken to preserve Criteo intellectual property; and how it can be leveraged to assess offline performance close to that of real-world production advertising systems.
The dataset is available in Hugging Face: \url{https://huggingface.co/datasets/criteo/CriteoPrivateAd}.

\subsection{Description}

\noindent \textbf{Overview.} The \texttt{CriteoPrivateAds} dataset contains 100 millions displays spanning 30 consecutive days of data and more than 100 relevant features to learn common bidding models. 
Indeed, we are providing binary labels such as click or landed click; as well as integer labels such as the number of sales attributed to a specific display.
Such labels are retrieved from third-party cookie traffic on Chrome and hence not necessarily representative of Chrome Protected Audiences traffic.
These choices have been made in order to have a sufficient number of positive examples enabling to learn relevant bidding models with a performance close to that of production systems.
Beyond features and labels relevant for training bidding models, we are also providing information required to simulate Privacy Sandbox APIs and emulate user-level DP, namely: hashed user ids, hashed publisher ids, hashed campaign ids (could be used as a proxy of advertiser ids) and time delta between the bid request timestamp and the click or sale event. This dataset is partionned by day to facilitate exploration, model seeding and train/validation/test split.\\

\noindent \textbf{Risk Mitigations.} In order to preserve Criteo intellectual property, we considered several privacy and business risk mitigations.
Regarding privacy risks, we anonymised feature values by hashing categorical ones, and via the use of a monotone transformation for continuous ones.
In addition, the major part of the features' names are not available.
Concerning business risks, labels in this dataset have been sampled non-uniformly from our production logs. Such labels could be sub-sampled from the released dataset to meet a specific business key performance indicator relevant for the advertising industry, e.g., a target click-through rate (CTR) representative of online traffic.
In \Cref{tab:dataset_overview2}, we are for instance providing examples of CTR that are common in the advertising industry.
In addition, some features used in production have been removed from this dataset without compromising its relevance to estimate bidding model performance.\\

\noindent \textbf{Reproducing Production Offline Performance.} 
We decided to sub-sample negative examples to ensure that the number of clicks remains sufficient for achieving representative performances, leading to a click-through rate CTR $\approx 0.36$ and a conversion rate CVR $\approx 0.02$, where a conversion here means a positive sale event (i.e., at least one product has been bought by the user). 
Hence, it is important to rescale validation metrics for interpretable results, close to production performance. 
Let $p_0$ be the CTR in \texttt{CriteoPrivateAds}, $p^\star$ the targeted CTR representative of online performance. 
By denoting $y\in \{0,1\}$ be the true label $y$,  $\hat{y} \in [0,1]$ the prediction  and $\ell$ the loss, we must use for validation the re-weighted loss \citep[Theorem 1]{10.5555/1642194.1642224}:
\begin{align}
        \tilde{\ell}(y,\hat{y}) =  y \ell(y,\hat{y}) + \frac{p_0(1-p^\star)}{p^\star(1-p_0)} 
        (1-y) \ell(y,\hat{y}) \ .
    \label{eq:1}
\end{align} 
While the re-scaling formula in \eqref{eq:1} is relatively simple, its proof based on Bayes' theorem is not as straightforward. 
We defer the interested reader to the seminal paper of \citet{10.5555/1642194.1642224} for the proof.
Note that a similar rescaling could be performed for other binary labels (e.g., sale or landed click event).

\subsection{Column Semantics}

Features used to train bidding models are grouped into five buckets with respect to their logging and inference constraints in Chrome Protected Audiences API. 
We chose the latter API as a reference for building this dataset as it stands for the most mature API to train bidding models without third-party cookies.
Note that this feature bucketisation into five groups can be easily relaxed to other browser vendors proposals or private learning frameworks as we are providing the precise semantics of such bucketisation.
Each row in the dataset stands for a specific display that has been made to a user. 
We are providing first some identifiers for each display in order to test many model training configurations, notably based on browser vendors proposals:

\begin{itemize}
    \item \textbf{id}: the id of the row.
    \item \textbf{uuid}: user id consistent over the day. 
    The same user will have two different user ids for two different days. 
    To clarify the difference between uuid and a device id, note that the data in \texttt{CriteoPrivateAds} comes mainly from the third-party cookie in the Chrome instance, and the rest from aggregated data from the 3rd-party cookie and app data from the same device (cookie id is matched with gaid from apps).
    \item \textbf{campaign\_id}: the id of the advertising campaign associated to the display, which could be used as a proxy for the advertiser id.
    \item \textbf{publisher\_id}: the id of the publisher on which the display has been made.
    \item \textbf{display\_order}: order of the display for the same user $\times$ campaign $\times$ day.
\end{itemize}

As highlighted previously, bidding model features are grouped into the following five buckets.
Albeit aligned with Protected Audiences API setting of Chrome Privacy Sandbox, this bucketisation is also coherent with what we could expect from a future open web without direct access to cross-domain user features via third-party cookies.
Indeed, the below features' groups allow to disentangle single-domain user features (also called first-party), cross-domain user features (e.g., publisher $\times$ advertiser or cross-advertiser), cross-domain features associated to a pre-defined user cohort (called Interest Group in Privacy Sandbox context), ad features (e.g., related to the ad campaign) and contextual features. 
These five buckets are detailed in what follows:

\begin{itemize}
    \item \textbf{features\_kv\_bits\_constrained} ($\approx 30$ different features): single-domain (a domain refers to a website) user features. 
    In Protected Audiences API of Chrome Privacy Sandbox, these features can be encoded in the \texttt{modelingSignals} field, subjected to the 12-bit constraint and available in the key-value server at inference time.
    \item \textbf{features\_kv\_bits\_not\_constrained} ($\approx 10$ different features): all the features derived from Interest Group (IG) name / renderURL, that is all ad features available in \texttt{reportWin} field outside of \texttt{modelingSignals} field in Protected Audiences API setting. These features are available in the key-value server at inference time and do not have any logging constraint (e.g., 12-bit encoding).
    \item \textbf{features\_browser\_bits\_constrained} ($\approx 10$ different features): all cross-advertiser user features available in \texttt{generateBid} in Protected Audiences API setting. This includes features that can be encoded in \texttt{recency} and \texttt{joinCount} fields. These features can be logged in \texttt{modelingSignals} field but are only available in the browser at inference time.
    \item \textbf{features\_ctx\_not\_constrained} ($\approx 10$ different features): all features only available in the contextual call with no logging constraints. These features stand for contextual ones.    \item \textbf{features\_not\_available} ($\approx 80$ different features): all cross-domain user features that cannot be used to train bidding models in the Privacy Sandbox context.
\end{itemize}

The labels available for training bidding models are:

\begin{itemize}
    \item \textbf{is\_clicked}: binary label indicating whether the display has been clicked or not.
    \item \textbf{is\_click\_landed}: binary label indicating whether the click has been observed on the advertiser website.
    \item \textbf{is\_visit}: binary label indicating whether the user interacted with the advertiser website after a landed click (at least one advertiser event after the landing event).
    \item \textbf{nb\_sales}: number of sales attributed to the clicked display. 
    \item The \textbf{is\_sale} label has to be created with \textbf{binarise(is\_visit * nb\_sales)}.
\end{itemize}

Given the aforementioned labels, this dataset enables the training and evaluation of models such as $\mathbb{P}$(click $\mid$ display), $\mathbb{P}$(landed\_click $\mid$ display), $\mathbb{P}$(sales $\mid$ landed\_click) and $\mathbb{E}$(number of sales $\mid$ landed\_click).
Note that $\mathbb{P}$(sales $\mid$ display) can also be obtained by using $\mathbb{P}$(sales $\mid$ landed\_click) x $\mathbb{P}$(landed\_click $\mid$ display) using the chain rule and under some independence assumptions.\\

Finally, we are also providing information about click and conversion delays with respect to the bid request timestamp in order to emulate the way Privacy Sandbox APIs are working, notably the Aggregated Reporting Attribution API and its event-level reporting based on the configured reporting window:

\begin{itemize}
    \item \textbf{click\_delay\_after\_display\_array}: : delay in minutes between click and display. If the click occurs the next day, it will still be reported in this field. 
    \item \textbf{landed\_click\_delay\_after\_display\_array}: delay in minutes between landed click and display. If the click occurs the next day, it will still be reported in this field.
    \item \textbf{sale\_delay\_after\_display\_array}: array of delays in minutes for multiple sales.  If the sales attributed to the display occur within the next 28 hours, they will be reported in this field. As an example, if there is a 7 a.m. display and a 7 a.m. conversion on the next day (24 hour delta), this field will be [24 $\times$ 3600 seconds]. Note that this array captures requests with multiple or aggregate labels (e.g., multiple sales) associated to the same display.
\end{itemize}

\subsection{Baseline experiments}
The \texttt{CriteoPrivateAds} dataset provides sufficient number of features and displays to learn an algorithm with relevant offline performance, ensuring its relevance for industrial benchmarks and applications.
In \Cref{tab:dataset_overview2}, we provide results obtained on this dataset using baseline models based on logistic regression, with a given targeted CTR and the associated re-weighted loss, as defined in \eqref{eq:1}.\\

\noindent \textbf{Metrics.} We used a rescaled formula of the log-likelihood (LLH), that is defined as $ \operatorname{LLHCompVN(f)}=\frac{\operatorname{LLH}(f_0) -\operatorname{LLH}({f})}{\operatorname{LLH}(f_0)}$  where $f$ is the trained bidding model and $f_0$ is the naive model always predicting the average label in the training dataset.
Another metric of interest is the calibration, defined as the ratio between the sum of predictions and the sum of the validation labels.
We would like to emphasise that classical classification metrics such as the area under the curve (AUC) are less relevant to benchmark bidding models since we need first and foremost calibrated models.
Note that our baseline results might be difficult to achieve because of the anonymisation and hashing of the columns and values.

\begin{table}[ht]
\centering
\renewcommand{\arraystretch}{1.2} 
\footnotesize
\begin{tabular}{|>{\raggedright}p{4cm}|p{1cm}|p{1cm}|p{1cm}|}
\hline
\diagbox{\textbf{Task}}{\textbf{Target CTR}} & \textbf{0,1\%} & \textbf{0,5\%} & \textbf{1\%} \\ \hline
Landed Click $\mid$ Display              &       0.170     &      0.186      &      0.234     \\ \hline
Sales $\mid$ Landed Click   &      0.218      &      0.218   & 0.218   \\ \hline
Sales $\mid$ Display   &      0.171      &      0.187   & 0.237   \\ \hline

\end{tabular}
\caption{Rescaled LLH of baseline bidding models for a given target CTR and a given task. Learned from day 1 to 25, validated from day 26 to 30.}
\label{tab:dataset_overview2}
\end{table}

\section{Private Bidding Optimisation Paradigms}
\label{sec:use-cases}

To ease the use of \texttt{CriteoPrivateAds}, we are providing in this section some brief private training paradigms to learn bidding models, that are currently considered or researched in the advertising industry.
As before, these use-cases are mainly inspired from the Chrome Privacy Sandbox but some of them are also aligned with where the industry is leaning.
Note that these use-cases are obviously not exhaustive and that \texttt{CriteoPrivateAds} is versatile enough to be used in many applications.

\subsection{User vs Display-level DP}

This dataset being sampled at the display level, it can be naturally used to design and test private bidding model training strategies based on display-level differential privacy (also coined item-level DP in the academic literature).
Albeit we are providing a user identifier, this dataset cannot be used directly, without involving some bias, to assess the relevancy of user-level DP frameworks.
Indeed, since we sampled on a display basis and not on a user one, the distribution of user's contributions in \texttt{CriteoPrivateAds} is different from the online traffic one.
To tackle this issue, we are providing two ways to use this dataset in a user-level DP context.\\

\noindent \textbf{\texttt{CriteoPrivateAds} might be representative to what is sent to a private aggregation server.} In the Privacy Sandbox context, the report collector (e.g., the adtech company willing to collect event-level reports to train bidding models) will have access to display-level reports (either encrypted or not), sampling per display could be performed by the report collector, which will then send the sampled batch of reports to a private aggregation server (e.g., a Private Aggregation service). This server will then implement a user-level DP framework, by for instance capping the number of displays per user (with some fixed or flexible logic, e.g., by sampling uniformly all user’s displays). The final dataset will then be used for aggregation or training purposes. 
To avoid any bias issue between offline training and online inference, we will need the sampling ratio of negatives and positives to align the training loss with the real data distribution.\\ 

\begin{figure}
    \centering
    \includegraphics[width=1\linewidth]{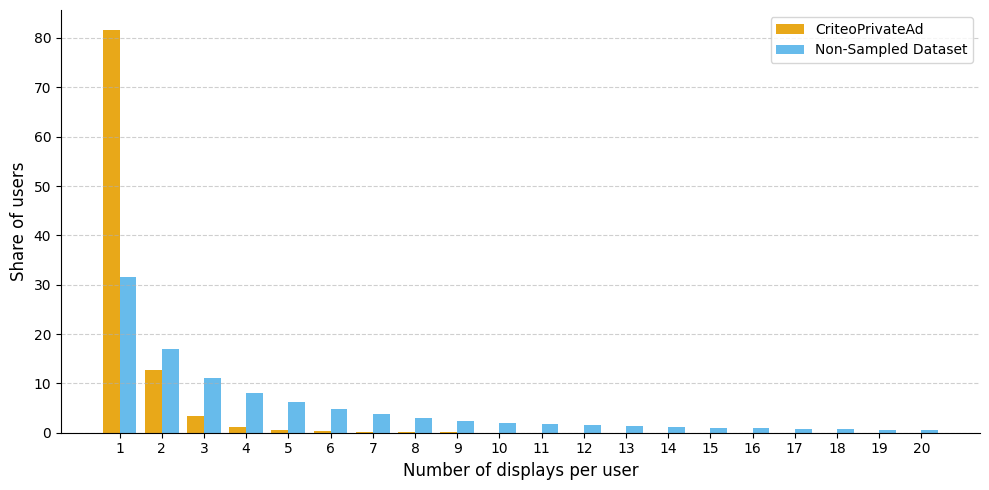}
    \caption{Discrepancy between empirical distributions of users' contributions in \texttt{CriteoPrivateAds} and associated online traffic.}
    \label{fig:user-dp}
\end{figure}

\noindent \textbf{\texttt{CriteoPrivateAds} could be aligned to the true user distribution.} In addition to \texttt{CriteoPrivateAds} dataset, we will provide the data used to plot the two histograms depicted in \Cref{fig:user-dp}.
To simulate what happens on the original non-sampled dataset, we propose a method to build a semi-synthetic dataset from \texttt{CriteoPrivateAds}, where the distribution of displays per user matches the original non-sampled dataset. 
By denoting $p$ the real distribution (depicted in blue in the above figure) and $q$ the approximate one (depicted in orange), we can build importance ratios $p/q$ per bucket and use them to sub-sample \texttt{CriteoPrivateAds} data to match $p$.
Naive sampling might lead to duplicated (if we fix a too large dataset size) or very few rows (if we sub-sampled directly each bucket) in the final dataset.
To avoid this issue, we can take all displays associated to the last bucket in \texttt{CriteoPrivateAds} (e.g., number of displays per user = 20 if the support is [1,20]), use $p$ to determine a dataset size (i.e., dataset size = nb\_users\_having\_20\_displays divided by Proba(nb\_displays\_per\_user = 20); and sample \texttt{CriteoPrivateAds} based on $p$.

\subsection{Protected Audiences API}

\begin{figure}[!ht]
\centering
\resizebox{0.5\textwidth}{!}{%
\begin{circuitikz}
\tikzstyle{every node}=[font=\normalsize]
\draw  (2.5,14) rectangle (6.25,12.75);
\node [font=\footnotesize, rotate around={21:(0,0)}] at (7.4,11.5) {contextual score $c$};
\draw  (2.5,11.5) rectangle (6.25,10.25);
\draw  (8.75,12.75) rectangle (12.5,11.5);
\node [font=\LARGE] at (5,13.25) {};
\node [font=\LARGE] at (5,9.75) {};
\draw [->, >=Stealth] (6.25,13.5) .. controls (7.5,13) and (7.5,13) .. (8.75,12.5) ;
\draw [->, >=Stealth] (12.5,12.1) -- (15,12.1);
\node [font=\large] at (4.3,13.5) {      KV Server};
\node [font=\large] at (4,11) {   \quad       Contextual call};
\node [font=\large] at (10.5,12.3) {In browser};
\node [font=\small] at (4.3,13) {User History};
\node [font=\small] at (4.3,10.5) {publisher / SSP};
\node [font=\small] at (10.5,11.87) {$\sigma(u+c)$};
\draw [->, >=Stealth] (6.25,10.75) -- (8.75,11.75);
\node [font=\footnotesize, rotate around={-23:(0,0)}] at (7.8,13.55) {user-campaign score $u$};
\node [font=\footnotesize] at (13.7,12.35) {Prediction};
\end{circuitikz}
}%
\caption{Two-tower inference architecture constraining how private bidding models are built.}

\label{fig:score_schema}
\end{figure}
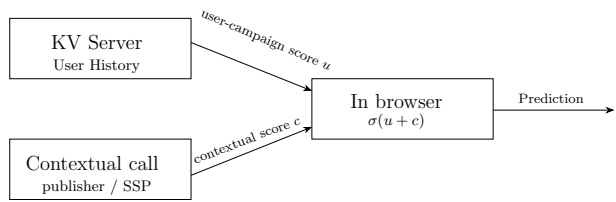

Learning using the current version of the Chrome Protected Audiences API is subjected to three main constraints that we list below in descending order of impact on learning metrics:

\begin{enumerate}
    \item \textbf{Two-tower inference paradigm.} The display log constructed with winning auctions is constrained by the inference system design. User history features can be converted to a user score in the key-value server while the contextual score is computed during the contextual call. 
    Then, it is not possible to model interaction between context/user features and campaign features. However, it is possible to compute a user-campaign score, and a contextual score, and gather them in browser.
    This two-tower inference architecture is depicted in \Cref{fig:score_schema}.
    \item \textbf{User Features' Quantisation.} The user features have to be stored in a 12-bit signal, coined \texttt{modelingSignals}, constraining the amount of information about user history that can be used by the bidding model.
    \item \textbf{Local DP.} User features are subjected to local differential privacy, via the randomised response mechanism defined in \Cref{def:random-response} on the complete 12-bit signal (not bit by bit).
    While event-level DP is expensive in general compared to global DP, it is currently negligible in Protected Audiences API setting as $\varepsilon \approx 12.9$, that is to say the randomised response provides a wrong user features' vector with probability $p=0.01$.
\end{enumerate}

\subsection{Aggregation API - Learning from Label Proportions}

Assuming that a feature vector $x$ lies in $\mathcal{X}$, and that labels are binary, i.e., $y \in \{0,1\}$, given $\mathbf{A} \in \mathcal{X}$ the Private Aggregation Service of Chrome Privacy Sandbox allows to get a noisy estimate of $\mathbb{P} \left ( y=1 | \mathbf{A} \right )$, by defining the key $k(x,\mathbf{A})=1$ if and only if $x \in \mathbf{A}$.
In other words, by using this Aggregation Service, we get noisy label proportions, where the noise is applied on the label proportion $\mathbb{P} \left ( y=1 | \mathbf{A} \right )$ using global differential privacy.
As of today, the Laplace mechanism is used in Chrome Privacy Sandbox, see \Cref{def:laplace-dp}.
The set $\mathbf{A}$ stands for the so-called "bag" in the learning from label proportions paradigm \citep{quadrianto2008estimating}.
This paradigm can be simulated using \texttt{CriteoPrivateAds} by aggregating display-level data and noising the associated label proportions.

\subsection{Private Model Training via DP-SGD}

Another private bidding model training framework we could consider using this dataset is the one akin to differentially private gradient descent, where gradients are noised using global differential privacy during training.
A seminal private training approach to meet this objective is DP-SGD \citep{Abadi_2016} and associated variants.

\section{Conclusion}

We release in an open-source manner \texttt{CriteoPrivateAds}, the largest and richest bidding dataset to learn a variety of click and conversion models.
This dataset can be used to learn standard bidding models but its structure has been designed to fit browser vendors private advertising proposals and associated constraints.
Through this open-source release, we hope to foster innovation in both the academic and industrial communities to design scalable private bidding optimisation strategies which help protecting users' data while ensuring an economically viable open internet.

\bibliographystyle{plainnat}  
\bibliography{references}  

\end{document}